\documentclass[11pt]{article}
\usepackage[dvips]{graphics}
\usepackage{latexsym} 

\def\submission{0} 

\newlength{\saveparindent}
\newlength{\saveparskip}
\newtheorem{thm}{Theorem}[section]
\newtheorem{lem}[thm]{Lemma}
\newtheorem{cor}[thm]{Corollary}
\newtheorem{propo}[thm]{Proposition}
\newtheorem{defn}[thm]{Definition}
\newtheorem{assm}[thm]{Assumption}
\newtheorem{clm}[thm]{Claim}
\newtheorem{rem}[thm]{Remark}
\newtheorem{fct}[thm]{Fact}
\newtheorem{egs}[thm]{Example}

\newenvironment{theorem}{\begin{thm}\begin{sl}}%
{\end{sl}\end{thm}}
\newenvironment{lemma}{\begin{lem}\begin{rm}}%
{\end{rm}\end{lem}}
{\end{sl}\end{cor}}
{\end{rm}\end{propo}}
{\end{rm}\end{defn}}
{\end{em}\end{assm}}
{\end{rm}\end{clm}}
{\end{rm}\end{rem}}
{\end{rm}\end{fct}}
{\end{rm}\end{egs}}

\newcommand{\secref}[1]{\mbox{Section~\ref{#1}}}
\newcommand{\apref}[1]{\mbox{Appendix~\ref{#1}}}
\newcommand{\thref}[1]{\mbox{Theorem~\ref{#1}}}

\newcommand{\lemref}[1]{\mbox{Lemma~\ref{#1}}}

\newcommand{\figref}[1]{\mbox{Figure~\ref{#1}}}
\newcommand{\eqref}[1]{\mbox{Equation~(\ref{#1})}}

\newcommand{\maybespace}{{\ifnum\springer=0{{\ }}\fi}}

\newcount\proofqeded
\newcount\proofended
\def\qed{{\hspace{3pt}\rule[-1pt]{3pt}{9pt}}
\end{rm}\addtolength{\parskip}{-0pt}
\setlength{\parindent}{\saveparindent}
\global\advance\proofqeded by 1 }
\newenvironment{proof}%
 {\proofstart}%
 {\ifnum\proofqeded=\proofended\qed\fi \global\advance\proofended by 1
  \medskip}
 {\proofsketchstart}%
 {\ifnum\proofqeded=\proofended\qed\fi \global\advance\proofended by 1
  \medskip}
\makeatletter
\def\proofstart{\@ifnextchar[{\@oprf}{\@nprf}}
\def\proofsketchstart{\@ifnextchar[{\@osprf}{\@nsprf}}
\def\@oprf[#1]{\begin{rm}\protect\vspace{6pt}\noindent{\bf Proof of #1:\ }%
\addtolength{\parskip}{5pt}\setlength{\parindent}{0pt}}
\def\@osprf[#1]{\begin{rm}\protect\vspace{6pt}\noindent{\bf Sketch of
Proof of #1:\ }
\addtolength{\parskip}{5pt}\setlength{\parindent}{0pt}}
\def\@nprf{\begin{rm}\protect\vspace{6pt}\noindent{\bf Proof:\ }%
\addtolength{\parskip}{5pt}\setlength{\parindent}{0pt}}
\def\@nsprf{\begin{rm}\protect\vspace{6pt}\noindent{\bf Proof Sketch:\ }%
\addtolength{\parskip}{5pt}\setlength{\parindent}{0pt}}

\newcounter{ctr}

\newcounter{ectr}

\newenvironment{centerlist}{%
\begin{list}{\mbox{}}{\labelwidth=0pt%
\labelsep=0pt \leftmargin=0pt \topsep=1pt%
\setlength{\listparindent}{\saveparindent}%
\setlength{\parsep}{\saveparskip}%
\setlength{\itemsep}{1pt} }}{\end{list}}

\newlength{\savejot}
\newenvironment{newmath}{\begin{displaymath}%
\setlength{\abovedisplayskip}{4pt}%
\setlength{\belowdisplayskip}{4pt}%
\setlength{\abovedisplayshortskip}{6pt}%
\setlength{\belowdisplayshortskip}{6pt} }{\end{displaymath}}

\newenvironment{newequation}{\begin{equation}%
\setlength{\abovedisplayskip}{4pt}%
\setlength{\belowdisplayskip}{4pt}%
\setlength{\abovedisplayshortskip}{6pt}%
\setlength{\belowdisplayshortskip}{6pt} }{\end{equation}}

\newcommand{\headingg}[1]{{\sc{#1}}}

\newcommand{\heading}[1]{{\vspace{5pt}\noindent\sc{#1}}}

\def\bits{\{0,1\}}

\def\cross{\times}

\newcommand{\mystrut}{\rule{0em}{15pt}}

\newcommand{\Moves}{\mathrm{Mvs}}
\newcommand{\EvenMoves}{\mathrm{EvMvs}}
\newcommand{\OddMoves}{\mathrm{OdMvs}}

\newcommand{\eqdef}{\stackrel{\rm def}{=}}

\def\suchthatt{\: :\:}
\def\next{\:;\:}
\newcommand{\set}[2]{\{\:#1 \suchthatt #2\:\}}
\newcommand{\sset}[1]{\{\:#1\:\}}
\def\leqq{\;\leq\;}
\def\eqq{\;=\;}
\def\geqq{\;\geq\;}

\newcommand{\E}{{\mbox{\bf E}}}

\newcommand{\Prob}[1]{{\Pr[\,{#1}\,]}}
\newcommand{\probb}[2]{{\Pr}_{#1}[\,{#2}\,]}

\newcommand{\Probb}[2]{{{\Pr}_{#1}\left[\,{#2}\,\right]}}

\newcommand{\CondProbb}[3]{{\Pr}_{#1}\left[\:
#2\:\left|\right.\:#3\:\right]}

\newcommand{\Probexp}[2]{{\Pr}\left[\: #1 \suchthatt #2\:\right]}

\makeatletter

\def\appearsin#1{\gdef\@appearsin{#1}}

\def\maketitle{\par
 \begingroup
 \def\thefootnote{\fnsymbol{footnote}}
 \def\@makefnmark{\hbox 
 to 0pt{$^{\@thefnmark}$\hss}} 
 \if@twocolumn 
 \twocolumn[\@maketitle] 
 \else \newpage
 \global\@topnum\z@ \@maketitle \fi\thispagestyle{plain}\@thanks
 \endgroup
 \setcounter{footnote}{0}
 \let\maketitle\relax
 \let\@maketitle\relax
 \gdef\@thanks{}\gdef\@author{}\gdef\@title{}\gdef\@appearsin{}
          \let\thanks\relax}
\def\@maketitle{\newpage
 \noindent \@appearsin
 \vskip 1in \begin{center}
 {\LARGE \@title \par} \vskip 1.5em {\large \lineskip .5em
\begin{tabular}[t]{c}\@author
 \end{tabular}\par} 
 \vskip 1em {\normalsize \@date} \end{center}
 \par
 \vskip 1.5em} 
\def\abstract{\if@twocolumn
\section*{Abstract}
\else \small 
\begin{center}
{\bf Abstract\vspace{-.5em}\vspace{0pt}} 
\end{center}
\quotation 
\fi}
\def\endabstract{\if@twocolumn\else\endquotation\fi}
\mark{{}{}} 

\newif\ifshortconferences
\shortconferencesfalse

\def\ending#1{{\count1=#1\relax
\count2=\count1
\divide\count2 by 100
\multiply\count2 by 100
\advance\count1 by -\count2
\ifnum\count1=11
th%
\else \ifnum\count1=12
th%
\else \ifnum\count1=13
th%
\else
\count2=\count1
\divide\count1 by 10
\multiply\count1 by 10
\advance\count2 by -\count1
\ifnum\count2=1
st%
\else \ifnum\count2=2
nd%
\else \ifnum\count2=3
rd%
\else th%
\fi\fi\fi\fi\fi\fi
}}

\def\Proceedings{\ifshortconferences {\sl Proc.}\else {\sl Proceedings}\fi}

\newcounter{confnum}

\def\conf#1#2{%
\setcounter{confnum}{#2}%
\addtocounter{confnum}{-\csname #1zero\endcsname}%
\ifnum\value{confnum}=1%
\expandafter\ifx\csname #1One\endcsname\relax%
\Proceedings\ {\sl of the} \arabic{confnum}\ending{\value{confnum}}\ 
\csname #1name
\endcsname,\ \csname #1pub\endcsname,\ 19#2%
\else \csname #1One\endcsname\fi%
\else\ \Proceedings\ {\sl of the}
\arabic{confnum}\ending{\value{confnum}}\ \csname #1name\endcsname,\ %
\csname #1pub\endcsname,\ 19#2\fi}

\def\focs{\conf{FOCS}}

\newcommand{\istcs}[1]{\ifnum#1=
93{{\sl Proceedings of the Second Israel Symposium on Theory 
and Computing Systems\/}, IEEE, 1993}\else{\ifnum#1=
95{{\sl Proceedings of the Third Israel Symposium on Theory 
and Computing Systems\/}, IEEE, 1995}\else
This ISTCS not yet defined!
\fi}\fi}

\def\svconf#1#2{%
\csname #1name\endcsname~#2 {\sl Proceedings}, 
Lecture Notes in Computer Science Vol.~\csname #1vol\endcsname{#2},
\csname #1ed\endcsname{#2} ed., Springer-Verlag, 19#2}

\def\crypto{\svconf{CRYPTO}}

\def\CRYPTOvol#1{\ifnum#1=
84{196}\else{\ifnum#1=
85{218}\else{\ifnum#1=
86{263}\else{\ifnum#1=
87{293}\else{\ifnum#1=
88{403}\else{\ifnum#1=
89{435}\else{\ifnum#1=
90{537}\else{\ifnum#1=
91{576}\else{\ifnum#1=
92{740}\else{\ifnum#1=
93{773}\else{\ifnum#1=
94{839}\else{\ifnum#1=
95{963}\else{\ifnum#1=
96{1109}\else{\ifnum#1=
97{1294}\else{\ifnum#1=
98{1462}\else{\ifnum#1=
99{??}\fi}\fi}\fi}\fi}\fi}\fi}\fi}\fi}\fi}\fi}\fi}\fi}\fi}\fi}\fi}\fi}

\def\CRYPTOed#1{\ifnum#1=
84{R.~Blakely}\else{\ifnum#1=
85{H.~Williams}\else{\ifnum#1=
86{A.~Odlyzko}\else{\ifnum#1=
87{C.~Pomerance}\else{\ifnum#1=
88{S.~Goldwasser}\else{\ifnum#1=
89{G.~Brassard}\else{\ifnum#1=
90{A.~J.~Menezes and S.~Vanstone}\else{\ifnum#1=
91{J.~Feigenbaum}\else{\ifnum#1=
92{E.~Brickell}\else{\ifnum#1=
93{D.~Stinson}\else{\ifnum#1=
94{Y.~Desmedt}\else{\ifnum#1=
95{D.~Coppersmith}\else{\ifnum#1=
96{N.~Koblitz}\else{\ifnum#1=
97{B.~Kaliski}\else{\ifnum#1=
98{H.~Krawczyk}\else{\ifnum#1=
99{}\fi}\fi}\fi}\fi}\fi}\fi}\fi}\fi}\fi}\fi}\fi}\fi}\fi}\fi}\fi}\fi}

\def\eurocrypt{\svconf{EUROCRYPT}}

\def\EUROCRYPTvol#1{\ifnum#1=
84{209}\else{\ifnum#1=
85{219}\else{\ifnum#1=
86{??}\else{\ifnum#1=   
87{304}\else{\ifnum#1=
88{330}\else{\ifnum#1=
89{434}\else{\ifnum#1=
90{473}\else{\ifnum#1=
91{547}\else{\ifnum#1=
92{658}\else{\ifnum#1=
93{765}\else{\ifnum#1=
94{950}\else{\ifnum#1=
95{921}\else{\ifnum#1=
96{1070}\else{\ifnum#1=
97{1233}\else{\ifnum#1=
98{??}\else{\ifnum#1=
99{??}\fi}\fi}\fi}\fi}\fi}\fi}\fi}\fi}\fi}\fi}\fi}\fi}\fi}\fi}\fi}\fi}

\def\EUROCRYPTed#1{\ifnum#1=
84{T.~Beth}\else{\ifnum#1=
85{F.~Pichler}\else{\ifnum#1=
86{??}\else{\ifnum#1=    
87{D.~Chaum}\else{\ifnum#1=
88{C.~Gunther}\else{\ifnum#1=
89{J-J.~Quisquater, J.~Vandewille}\else{\ifnum#1=
90{I.~Damg{\aa}rd}\else{\ifnum#1=
91{D.~Davies}\else{\ifnum#1=
92{R.~Rueppel}\else{\ifnum#1=
93{T.~Helleseth}\else{\ifnum#1=
94{A.~De~Santis}\else{\ifnum#1=
95{L.~Guillou and J.~Quisquater}\else{\ifnum#1=
96{U.~Maurer}\else{\ifnum#1=
97{W.~Fumy}\else{\ifnum#1=
98{K.~Nyberg}\else{\ifnum#1=
99{??}\fi}\fi}\fi}\fi}\fi}\fi}\fi}\fi}\fi}\fi}\fi}\fi}\fi}\fi}\fi}\fi}

\def\AUSCRYPTvol#1{\ifnum#1=
92{718}\fi}
\def\AUSCRYPTed#1{\ifnum#1=
92{}\fi}

\def\asiacrypt{\svconf{ASIACRYPT}}

\def\ASIACRYPTvol#1{\ifnum#1=
91{739}\else{\ifnum#1=
94{917}\else{\ifnum#1=
96{1163}\else{??}\fi}\fi}\fi}

\def\ASIACRYPTed#1{\ifnum#1=
91{H.~Imai, R.~Rivest and T.~Matsumoto}\else{\ifnum#1=
94{J.~Pieprzyk and R.~Safavi-Naini}\else{\ifnum#1=
96{M.~Y.~Rhee and K.~Kim}\else{??}\fi}\fi}\fi}

\newcommand{\ignore}[1]{}

\def\Adv{\mathbf{Adv}}
\def\SuccIdeal{\mathbf{Succ}}
\def\SecIdeal{\mathbf{Sec}}

\def\DF{\df{F}}
\def\TF{\tf{F}}
\def\Const{\Opp{F}}
\def\Consto{\Oppo}
\def\DES{\mbox{{\rm DES}}}

\def\MD5{\mbox{MD5}}

\def\bits{\{0,1\}}
\def\Colon{:}

\newcommand{\remove}[1]{}

\newcommand{\BC}[2]{{\rm BC}(#1,#2)}  
\newcommand{\PERM}[1]{{\rm PERM}(#1)}  
\def\keylen{{\kappa}}
\def\blocklen{{n}}
\def\evand{\:\wedge\:}
\def\concat{\|}

\def\bad{\textsc{bad}}
\def\notbad{\overline{\textsc{bad}}}

\def\F{F}
\def\E{E}
\newcommand{\ideal}[1]{\texttt{Ideal}(#1)}
\newcommand{\triple}[1]{\texttt{Triple}_{#1}}

\newcommand{\psDouble}[1]{\texttt{psDouble}(#1)}

\newcommand{\df}[1]{\mathsf{Dbl}{\mbox{-}#1}}
\newcommand{\dfo}{\mathsf{Dbl}}
\newcommand{\Opp}[1]{\mathsf{Op}{\mbox{-}#1}}
\newcommand{\Oppo}{\mathsf{Op}}
\newcommand{\tf}[1]{\mathsf{Trp^2}{\mbox{-}#1}}
\newcommand{\tfo}{\mathsf{Trp^2}}

\def\vview{\textrm{View}}

\def\rvview{\textsf{View}}

\def\rvk{\textsf{k}}

\def\rvE{E}
\def\rvF{F}
\def\rvQ{\textsf{Q}}
\def\rvR{\textsf{R}}
\def\rvT{\textsf{T}}
\def\rvU{\textsf{U}}

\def\SeenKeyPairs{\mathrm{SKP}}
\def\RemKeyPairs{\mathrm{RKP}}
\def\NewKeyPairs{\mathrm{NKP}}
\def\Equeries{\mathrm{E}\mbox{-}\mathrm{Qrs}}
\def\Fqueries{\mathrm{F}\mbox{-}\mathrm{Qrs}}

\begin{document}


\def\ourtitle{{\LARGE\bf 
Security Amplification by Composition: The case of
Doubly-Iterated, Ideal Ciphers}}

\def\ourdate{June 1998}

\def\ourabstract{We investigate, in the Shannon model, the security of
  constructions corresponding to double and (two-key) triple DES. That
  is, we consider $F_{k_1}(F_{k_2}(\cdot))$ and
  $F_{k_1}(F_{k_2}^{-1}(F_{k_1}(\cdot)))$ with the component
  functions being ideal ciphers. This models the resistance of these
  constructions to ``generic'' attacks like meet in the middle attacks.
  
  We obtain the first proof that composition actually
  increases the security in some meaningful sense.  We compute a bound
  on the probability of breaking the double cipher as a function of
  the number of computations of the base cipher made, and the number
  of examples of the composed cipher seen, and show that the success
  probability is the square of that for a single key cipher. The
  same bound holds for the two-key triple cipher. The first bound
  is tight and shows that meet in the middle is the best possible
  generic attack against the double cipher.}

\def\ourkeywords{ 
\noindent {\bf Keywords:} Ciphers, cascaded ciphers, Shannon model,
information theory, DES, Double DES, meet in the middle attacks.}

\ifnum\submission=1
\appearsin{Crypto 98 submission.}
\else
\appearsin{An extended abstract of this work appears in \crypto{98}.
This is the full version.}
\fi

\title{\ourtitle\vspace{0.3in}}

\author{
{\sc W.~Aiello}\thanks{\ Bellcore, 445 South St., Morristown, NJ 07960, USA.
E-Mail: \texttt{aiello@bellcore.com}}\and
{\sc M.~Bellare}\thanks{\ 
\ifnum\submission=1{\bf Contact Author.}\fi Dept.~of Computer Science \&
Engineering, University of California at San Diego, 9500 Gilman Drive, La
Jolla, CA~92093, USA.  E-Mail: \texttt{mihir@cs.\allowbreak ucsd.edu}.
\ifnum\submission=1 Phone: {619-534-4544}. FAX: {619-534-7029}.\fi URL: {\tt
http://\allowbreak www-cse.\allowbreak ucsd.edu/\allowbreak users/mihir}.
Supported in part by NSF CAREER Award CCR-9624439 and a 1996 Packard
Foundation
Fellowship in Science and Engineering.}
\and {\sc G.~Di Crescenzo}\thanks{\ 
Dept.~of Computer Science \& Engineering, University of California at San
Diego, 9500 Gilman Drive, La Jolla, CA~92093, USA.  E-Mail:
\texttt{giovanni@cs.\allowbreak ucsd.edu}. Supported in part by above
mentioned
grants of Bellare.}\and
{\sc R.~Venkatesan}\thanks{\ Microsoft Research, One Microsoft Way,  
Redmond, WA 98052, USA. E-Mail: \texttt{venkie@microsoft.com}}
\vspace{0.3in}}

\date{\ifnum\submission=0\ourdate \\ \protect\vspace{0.2in}\fi}

\maketitle
\thispagestyle{empty}

\begin{abstract}
\ourabstract
\end{abstract}

\vspace{0.4in}
\ourkeywords

\newpage
\thispagestyle{empty}

\ifnum\submission=1


\noindent Crypto 98 submission.
\vspace{0.6in}

\begin{center}
\ourtitle
\end{center}

\vspace{0.4in}

\begin{center}
{\bf Abstract} \\[15pt]
\begin{small}\begin{minipage}{5.5in}
\setlength{\parindent}{\saveparindent}
\ourabstract
\end{minipage}\end{small}
\end{center}

\vspace{0.4in}
\ourkeywords

\fi

\ifnum\submission=0
\tableofcontents
\newpage
\else
\mbox{ }\newpage
\setcounter{page}{1}
\fi


\section{Introduction}

A block cipher is a map $F\Colon\bits^\keylen\cross\bits^{\blocklen}\to
\bits^{\blocklen}$.  Here $\keylen$ is the key size and $\blocklen$ is the
block size. Each $\keylen$-bit key $k$ induces a map $F_k(\cdot)\eqdef
F(k,\cdot)\Colon \bits^{\blocklen}\to\bits^{\blocklen}$ which is a
\textit{permutation\/} on $\bits^{\blocklen}$. Let $F^{-1}$ denote the inverse
cipher, meaning $F^{-1}(k,\cdot)\eqdef F_k^{-1}$ is the inverse map of
$F_k(\cdot)$. For example, $\DES$ is such a cipher with $\keylen=56$ and
$\blocklen=64$.

It is common practice to compose ciphers in attempts to increase security.
The
result of composition is a new cipher, with a larger key size but the same
block size. Here are the two most popular mechanisms, corresponding,
respectively, to double DES and (two-key) triple DES:
\begin{itemize}
\item {\em Double $F$, or the 2-cascade cipher:} $\DF\Colon\bits^{2\keylen}
\cross\bits^{\blocklen}\to\bits^{\blocklen}$ is defined by
\begin{newmath}
\DF_{k_1,k_2}(x)\eqq F_{k_1}(F_{k_2}(x)) \;.
\end{newmath}%
\item {\em Two-key triple $F$:} $\TF\Colon\bits^{2\keylen}\cross
\bits^{\blocklen}\to\bits^{\blocklen}$ is defined by
\begin{newmath}
 \TF_{k_1,k_2}(x) \eqq F_{k_1}(F_{k_2}^{-1}(F_{k_1}(x))) \;.
\end{newmath}%
\end{itemize}
Let
$\Const\Colon\bits^{\keylen^*}\times\bits^{\blocklen}\to\bits^{\blocklen}$
denote one of these, where $\keylen^*=2\keylen$ and $\Oppo\in\{\dfo,\tfo\}$.
What we want to know is: How good a cipher is $\Const$?  Has the composition
and the increased key length actually bought us anything?

\heading{Generic versus cryptanalytic attacks.} There are several possible 
approaches to this question, depending on what kinds of attacks one wants to
take into account. There are two main classes of attacks:
\begin{itemize}
\item {\em Cryptanalytic attacks:\/} Like differential \cite{bs1,bs2} and 
linear \cite{matsui} cryptanalysis
\item {\em Generic attacks:\/} Like exhaustive key search and
meet-in-the-middle
attacks.
\end{itemize}
Generic attacks are, roughly, those that don't exploit the structure of the
cipher, but work against any cipher, even an ideal one. More precisely, we
define generic attacks as those that succeed in the Shannon model of an 
ideal cipher discussed below.

The strength of specific composed ciphers like double DES against cryptanalytic
attacks is not known; certainly, one does not expect a proof of such
strength. The strength of the composed cipher against generic attacks, in
contrast, can at least in principle be determined, by an analysis in the
Shannon model, since it is a purely information theoretic question. However,
the technical problems here are quite challenging; in particular, it is not
even known that composition increases the strength of a cipher at all in this
model.

In this paper we tackle this question, analyzing, in the Shannon model, two-key
based compositions such as the above.  We will prove upper bounds on the
probability of ``breaking'' the composed cipher as a function of the ``effort''
invested by the adversary, with both terms in quotes to be properly
defined. Our results are the first to show that cipher composition in the
Shannon model actually increases security: the success probability of an
adversary, as a function of her resources, is significantly lower than in the
case of a single key cipher. For the double cipher our results are actually
tight (optimal) and show that meet in the middle is the best possible generic
attack on this cipher.  We now define the model, and state our results, more
precisely.

\subsection{The model}

We model $F$ as an \textit{ideal} block cipher in the sense of Shannon. This
means $F(k,\cdot)$ is a \textit{random} permutation on $\bits^n$, for each
$k$. More precisely, let $\PERM{n}$ be the set of all permutations on
$\bits^n$.  Then, for each $\keylen$-bit key $k$, select, uniformly and
independently, a map {from} $\PERM{n}$, and assign $F_k$ this value.  So $F$
consists of $2^\keylen$ maps, each a random permutation.

Now, we want to ask how good is $\Consto$ as a composition operator.  How can
we measure this? We do so in a strong adversarial model, which allows the
adversary chosen plaintext attacks on $\Const$.  Furthermore, success for the
adversary $A$ does not mean she has to find the key: it suffices that $A$
identify some ``weakness'' in the cipher. This means $A$ should be able to
detect any deviation in $\Const_{k^*}(\cdot)$ {from} a truly random
permutation, when $k^*$ is a random and hidden key for $\Const$.

Formally, give the adversary oracles for $F,F^{-1}$.  (This models her ability
to compute the original cipher at any points she likes.)  Also give her an
oracle we call $E\Colon\bits^n\to\bits^n$, which can take one of two forms:
\begin{itemize}
\item {\em World 1:\/} Set $E=\Const_{k^*}(\cdot)$ where $k^*\in
\bits^{\keylen^*}$ is a randomly chosen key for cipher $\Const$ 
\item {\em World 2:\/} Set $E=\pi$ where $\pi$ is a permutation chosen 
randomly {from} $\PERM{n}$.
\end{itemize}
Put the adversary $A$ in one of these worlds, and ask her which one she is
in. If she can't tell then $\Const_{k^*}(\cdot)$ is behaving like a random
permutation, meaning it is good.  Formally, define the \textit{advantage\/} of
$A$ as $P_1-P_2$, where $P_i$ is the probability that $A$ outputs $1$ in world
$i\in \{1,2\}$.  (The probability is over the choice of the oracles in each
case.)  Call $A$ a \textit{$(q,t)$-adversary} if it makes at most $t$ queries
to the $F,F^{-1}$ oracles and at most $q$ queries to the $E$ oracle. (Note in
practice $t$ is likely to be much larger than $q$ since $F,F^{-1}$ queries are
just DES computations and $E$ queries are plaintexts in a chosen plaintext
attack.  We always assume $q \geq 1$ since otherwise the advantage of the
adversary is zero no matter what the construction.) Define
\begin{newmath}
   \SecIdeal(\Consto,\keylen,\blocklen,q,t) 
\end{newmath}%
as the maximum advantage attainable by any $(q,t)$-adversary. This is the key
quantity; it is a function we call the \textit{security} of the operator
$\Consto$.  The question is to determine this function as accurately as
possible. In particular we want to upper bound it as a function of the
adversary resources $q,t$ and the block cipher parameters $\keylen,n$.

Before stating the results we stress the power of the model. It allows chosen
plaintext attacks on the composite cipher $\Const$. Note it certainly captures
common attacks like birthday attacks and meet-in-the-middle attacks, but also
more sophisticated attacks which could be adaptive.

Notice that the advantage of a $(q,t)$ adversary in attacking the single key
cipher $F$ itself in this model (namely $E=F_{k}$ for a random $\keylen$ bit
string $k$ in world~1) will be (at most) $t/2^{\keylen}$.  This is the mark we
have to beat if we want to show that the composed cipher is stronger than the
original one.

\subsection{The results}

\begin{figure}[tb]
\centerline{\scalebox{0.7}{\includegraphics{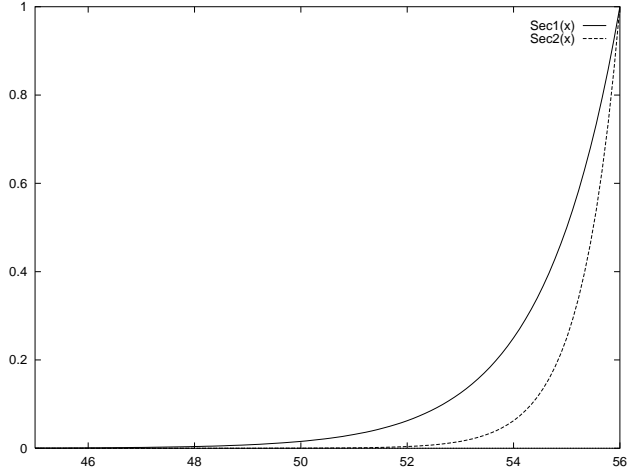}}}
\caption{{\small\sl ${\rm Sec1}(x)$ (the upper curve) and ${\rm Sec2}(x)$ (the
lower curve) are, respectively, the maximal possible advantage obtainable by an
adversary in breaking the single and double key ideal ciphers, respectively, as
a function of $x=\log_2(t)$, the logarithm of the number of cipher computations
made. We are using a key length of $\keylen=56$.  We see that ${\rm Sec2}$ lies
below ${\rm Sec1}$ but they meet at $1$. The text provides the exact formulas
for these quantities.}}\label{fig-plot}
\end{figure}

It is known that the strength of the composed cipher is at least that of the
first \cite{MaMa}, but prior to this work it was not known whether the
advantage of a $(q,t)$ adversary versus $\DF$ was any lower than its advantage
versus the single key cipher $F$ itself.  Here we are able to show that
composition actually increases security, in the ideal cipher model described
above.

\heading{The double key cipher.} Recall that the double $F$ cipher $\DF$ has
$2\keylen$ bits of key. Our main result is \thref{th-double-des}, which says
that $\SecIdeal(\Consto,\keylen, \blocklen,q,t)$ is at most $t^2/2^{2
\keylen}$. Namely, no $(q,t)$-adversary attacking the double cipher can achieve
an advantage greater than $t^2/2^{2\keylen}$.

We also show this bound is essentially tight, due to (a variant of) the meet in
the middle attack. \thref{th-att-double-des} presents an adversary who runs 
this attack, and analyzes it to show that its advantage is within a small
factor of $t^2/2^{2\keylen}$.

Note that the maximum possible advantage of an adversary attacking the double
cipher case is the square of the maximum possible advantage of an adversary of
the same resources attacking the original single key cipher. Thus, it is
considerably smaller in most cases.  (For example if $\keylen=56$ and
$t=2^{45}$ then the former is $2^{-22}$ and the latter is $2^{-11}$.  Or,
looking at it another way, to achieve an advantage of $2^{-11}$ against the
double cipher you need at least $2^{50}$ queries, while to get the same
advantage against the single cipher you need only $2^{45}$ queries.) To see the
relation better, we plot in \figref{fig-plot} the maximal advantage
$t/2^{\keylen}$ of an adversary in breaking the original single key cipher, and
the maximal advantage $t^2/2^{2\keylen}$ of an adversary in breaking the double
cipher, as a function of $x=\log_2(t)$.

Notice that the upper bound on the advantage in the double key case hits one
(meaning, the scheme can be broken) when $t=2^{\keylen}$. This is expected:
that's the meet in the middle attack.  Of course, that's the same point at
which the advantage hits one for the original single key cipher. (In this case
due to an exhaustive key search attack.) Thus, the ``effective key length'' of
the double cipher is not more than that of the single one.  That does not mean
that security has not increased. Security is not a number, but a function of
the resources invested, and our analysis and \figref{fig-plot} show that
for values of $t$ below $2^{\kappa}$ the chance of breaking the double
cipher is smaller than that of breaking the original one.

\heading{The two-key triple cipher.} We show that the same bound holds for the
two-key triple cipher, meaning the advantage of a $(q,t)$ adversary is bounded
by $t^2/2^{2\keylen}$. This shows that here too there is an improvement in the
security curve as a function of $t$. In this case our bound is tight for the
case $t\approx q$ but not tight in general.

\heading{The $m$-fold cascade.} The $m$-fold composition of cipher $F$ is the
cipher with key $k_1,\ldots,k_m$ defined by $F_{k_1,\ldots,k_m} = F_{k_1} \circ
F_{k_2}\circ\cdots\circ F_{k_m}$. The techniques above extend to show that the
advantage of an $(q,t)$ adversary is at most $t^m/2^{m\keylen}$.  This shows
that the advantage grows more and more slowly as $m$ increases. However, for $m
\geq 3$ the result is not tight; we expect the $3$-fold composed cipher to have
an even greater strength than this indicates. Thus, we won't discuss this
result any more in this paper.

\heading{The future.} The analysis of the two key ciphers we present
here is a start on a problem that appears to be quite technically
challenging.  In the future we would like to see tight bounds on the
advantage for the $m$-fold composition for $m\geq 3$ and also for the
two-key triple cipher in the case $q<<t$, but the distance needed to
get there seems quite large at this time.

\subsection{Related work}

The model used here is that of Kilian and Rogaway \cite{KiRo}, who in turn
built on Even and Mansour \cite{EvMa}, although the basic idea of course
goes
back to Shannon \cite{Shannon}.

Kilian and Rogaway \cite{KiRo} analyze Rivest's DESX cipher in this model and
show it has a large effective key length.  If generic (or, as they call them,
key search) attacks are the only concern, DESX is cheaper than Double or Triple
DES, but DESX is just as vulnerable as DES to differential and linear
cryptanalysis. The (apparent) strength of Double and two-key triple DES against
cryptanalysis coupled with the proven strength against generic attacks seem to
make a strong combination that is absent for DESX.

The basic meet in the middle attacks are due to \cite{DiHe,MeHe}.  Even and
Goldreich provide some time-space tradeoffs for meet-in-the-middle attacks
\cite{EvGo}, and Van Oorschot and Wiener \cite{vw} reduce the space
requirements.

Even and Goldreich \cite{EvGo} had shown that the cascade of $m$ ciphers is at
least as strong as its strongest component. Maurer and Massey \cite{MaMa}
argued that this result required restrictions in the model, and also showed
that the cascade is at least as strong as its first component.  Our work is the
first to show that the cascade can be \textit{stronger} than the original
cipher.

Our analysis builds on techniques of \cite{KiRo} and \cite{bkr}.  Applications
aside, we feel that we are looking at a basic information theoretic question,
namely the power of cascaded ciphers.

A preliminary version of our paper appeard as \cite{abdv}. Material omitted
there due to space restrictions is included here.

\subsection{Discussion on Implications of our result}

What implications do these results have for the security of real ciphers like
DES? This is a question that needs to be addressed with some care. After all,
DES is not an ideal cipher.

We are not claiming to have ``proven Double DES'' secure; that
obviously is not a realistic possibility. Our results might be
interpreted as saying that the existence of a generic attack against
DES that is substantially better than the meet in the middle attack
would imply that there are serious weaknesses in the random behavior
of DES that so far has empirical support.
%

The class of generic attacks is broad enough to be interesting, including
meet-in-the-middle attacks and variants of it. But it does not include
cryptanalytic attacks like differential or linear cryptanalysis, which exploit
the structure of the cipher.  However, one should note that at the moment the
best attacks against Double and Triple DES are not the cryptanalytic ones, but
the generic meet-in-the-middle attacks. And our results can be interpreted as
ruling out improvements along those lines.

The adversary resources we consider here are the number of cipher computations
$t$ and the number of available plaintext-ciphertext pairs of the attacked
cipher available, $q$. These are the most basic resources, and also the natural
ones to consider in an information theoretic setting. One might attempt to
consider other resources like space (e.g.~when it is small compared to the
number of queries), or make a distinction between parallelizable and sequential
computations.  Addressing these issues would change the nature of the problem
to the point where it is difficult to see how it might be treated by techniques
similar to the ones we use.
%
%
%

\subsection{Organization}

The double cipher analysis is in \secref{se-double-des}. There we state and
prove the upper bound. In \apref{se-att-df} we present the meet in the middle
attack analysis that shows the upper bound is tight.  The analysis of the
two-key triple cipher is in \apref{se-triple}.

\section{Definitions}\label{sec-defs}

\headingg{General.} We use standard notation for expressing probabilistic
experiments and algorithms. Namely if $S$ is a probability space then $x\gets
S$ denotes the operation of drawing $x$ at random according to distribution
$S$. If $S$ is a set we use the same notation with the understanding that $S$
is imbued with the uniform distribution. If $S$ is not a set or probability
space (in particular if $x$ is a string or function) then $x\gets S$ is simply
an assignment statement.

\heading{Block ciphers.}  For an integer $\blocklen\geq 1$ let
$\PERM{\blocklen}$ denote the set of all maps $\pi\Colon\bits^{\blocklen}\to
\bits^{\blocklen}$ that are permutations, meaning both one-to-one and onto.  A
function $F\Colon \bits^{\keylen} \times \bits^{\blocklen}\to
\bits^{\blocklen}$ is a \emph{block cipher} if for each \emph{key} $k\in
\bits^{\keylen}$, the function $F(k,\cdot)\Colon\bits^{\blocklen}\to
\bits^{\blocklen}$ is a permutation on $\bits^n$, meaning a member of
$\PERM{\blocklen}$.  Here, $\blocklen$ is the block length of the cipher and
$\keylen$ is the key length of the cipher.  Think of $F$ as a $2^{\keylen}$ by
$2^{\blocklen}$ table, with entry $(k,x)$ containing $F(k,x)$. Each row is a
permutation of $\bits^n$.  For convenience, define $F_k \Colon
\bits^{\blocklen} \rightarrow \bits^{\blocklen} $, for each $k\in
\bits^{\keylen}$, by $F_k(x) = F(k,x)$.  This is the permutation in the $k$-th
row.  Although the function $F$ does not have an inverse function, it does have
a well defined inverse block cipher.  When it is clear {from} context that $F$
is a block cipher then we will let $F^{-1}\Colon \bits^{\keylen} \times
\bits^{\blocklen}\to \bits^{\blocklen}$ denote the block cipher inverse of $F$,
defined as follows: $F^{-1}(k,y) = F_k^{-1}(y)$.  That is, $F^{-1}(k,y) = x$
iff $F(k,x) = y$.

Let $\BC{\keylen}{\blocklen}$ denote the set of all block ciphers with key
length $\keylen$ and block length $\blocklen$. This is viewed as a
probability
space under the uniform distribution. Thus $F\gets \BC{\keylen}{\blocklen}$ 
means that $F$ is selected according to the following experiment:
\begin{tabbing}
123\=\kill
\>  \texttt{for all} $k\in\bits^{\keylen}$ \texttt{do} 
   $F(k,\cdot)\gets\PERM{n}$.
\end{tabbing}

\heading{Operators: Double and triple.}  We are interested in transformations,
or operators, which map one block cipher to another. In general such an
operator is a map $\Oppo$ taking a block cipher $F\in \BC{\keylen}{\blocklen}$
and returning another block cipher, which we denote by $\Opp{F}$, and which
belongs to $\BC{\keylen^*}{\blocklen^*}$ for some values of
$\keylen^*,\blocklen^*$ that depend on $\keylen,\blocklen$ and $\Oppo$. (In
this paper it will always be the case that $\blocklen^*=\blocklen$.) We now
define the two central operators for this paper.

The \textit{double composition operator} $\dfo\Colon\BC{\keylen} {\blocklen}\to
\BC{2\keylen}{\blocklen}$ is defined by $\df{F}_{k_1k_2}= F_{k_1}\circ
F_{k_2}$. In other words, $\df{F}(k_1k_2,x)=F(k_1,F(k_2,x))$ for every
$k_1,k_2\in\bits^{\keylen}$ and every $x\in\bits^{\blocklen}$. The \textit{two
key, triple composition operator} $\tfo\Colon\BC{\keylen} {\blocklen}\to
\BC{2\keylen}{\blocklen}$ is defined by $\tf{F}_{k_1k_2}= F_{k_1}\circ
F_{k_2}^{-1}\circ F_{k_1}$. In other words, $\tf{F}(k_1k_2,x)=
F(k_1,F^{-1}(k_2,F(k_1,x)))$ for every $k_1,k_2\in\bits^{\keylen}$ and every
$x\in\bits^{\blocklen}$. Note both these ciphers have key length twice that of
the original cipher.

\heading{Security.} We will be considering the security of these operators.
The setting for security is the following. Consider an adversary algorithm $A$
which has access to three oracles, $E,F,F^{-1}$, where $F\in
\BC{\keylen}{\blocklen}$ and $E\Colon \bits^{\blocklen}\to \bits^{\blocklen}$.
It computes with them and eventually outputs a bit. This computation is
\textit{adaptive}.  This means that it makes queries to oracles as it pleases,
choosing these queries as a function of answers to previous queries. We
represent $A$'s output when interacting with these oracles by
$A^{E,F,F^{-1}}$. (Since we will not restrict the computational power of the
adversary $A$, it is without loss of generality deterministic, and hence this
output is uniquely defined once $A,F,E$ are fixed.)  If the oracles that $A$
interacts with are chosen according to some distribution then $A$'s output
will be a random variable over $\bits$.  We let
\begin{newmath}
\SuccIdeal_A(\keylen,\blocklen) \eqq
\Probexp{A^{E,F,F^{-1}}=1}{F\gets\BC{\keylen}{\blocklen}\next
      E\gets\PERM{\blocklen}} 
\end{newmath}%
denote the success probability of $A$ in the ``ideal world'' (called world~2 in
the Introduction) where $E$ is a random permutation independent of the cipher
$F$. On the other hand, if
$\Oppo\Colon\BC{\keylen}{\blocklen}\to\BC{\keylen^*}{\blocklen^*}$ is an
operator then we let
\begin{newmath}
\SuccIdeal_A(\Oppo,\keylen,\blocklen) \eqq
\Probexp{A^{E,F,F^{-1}}=1}{F\gets\BC{\keylen}{\blocklen}\next
      k^*\gets \bits^{\keylen^*}\next E\gets \Opp{F}_{k^*}} \;.
\end{newmath}%
In other words, having selected $F$, apply the operator to it to get a new
cipher $F^* = \Opp{F}$. Now, choose at random a permutation $E$ of this
cipher, by choosing a key $k^*$ and setting $E$ to $F^*_{k^*}$. (This
was called world~1 in the Introduction.) Now let
\begin{newmath}
  \Adv_A(\Oppo,\keylen,\blocklen) \eqq 
  \SuccIdeal_A(\Oppo,\keylen,\blocklen) 
  \: - \: \SuccIdeal_A(\keylen,\blocklen) \;.
\end{newmath}%
This is the the advantage of $A$ in breaking the $\Oppo$ induced cipher.  To
measure the quality of a particular operator $\Oppo$ (eg.~$\dfo$ or $\tfo$) we
want to upper bound the advantage in terms of the resources used by the
adversary, meaning the number of queries it makes to its oracles.  We call a
query to the $E$ oracle an $\E$-query; a query to the $F$ oracle an $\F$ query;
a query to the $F^{-1}$ oracle an $\F^{-1}$ query. Typically the number of
$\E$-queries is denoted $q$, while the sum of the number of $\F$ and $\F^{-1}$
queries is denoted $t$. The security of the operator $\Oppo$ is then given by
\begin{newmath}
\SecIdeal(\Oppo,\keylen,\blocklen,q,t) \eqq \max_{A} 
   \Adv_A(\Oppo,\keylen,\blocklen) \;,
\end{newmath}%
where the maximum is taken over all adversaries $A$ who make at most $q$
$\E$-queries and at most $t$ $\F/\F^{-1}$ queries.  Thus our goal will be to
bound $\SecIdeal(\Oppo,\keylen,\blocklen,q,t)$ in terms of $q,t,\keylen,
\blocklen$ for the two ciphers we are investigating, namely $\Oppo=\dfo$ and
$\Oppo=\tfo$.

We stress that this bound will apply to any adversary. No assumptions are made
about the strategy followed by this adversary other than that it is limited to
the specified number of queries.

\section{Security analysis of the double cipher}
\label{se-double-des}

In this section our goal will be to determine the security of the doubly
iterated ideal cipher. In other words, we want to estimate, as accurately as
possible, the value of $\SecIdeal(\dfo,\keylen,\blocklen,q,t)$, as a function
of the cipher parameters $\keylen,\blocklen$ and the adversary resource bounds
$q,t$. The following is the main theorem, which provides an upper bound on the
security. It says that the advantage of any adversary $A$ attacking the doubly
iterated ideal cipher is at most $t^2/2^{2\keylen}$, regardless of the strategy
used by this adversary.

\begin{theorem}\label{th-double-des} For any $\keylen,\blocklen,q,t \geq 1$ it
is the case that
\begin{newmath}
 \SecIdeal(\dfo,\keylen,\blocklen,q,t)\leqq \frac{t^2}{2^{2\keylen}} \;.
\end{newmath}%
\end{theorem}
Notice that the bound depends only on the number $t$ of $F/F^{-1}$ queries made
by $A$, and the key length $\keylen$ of the cipher; it does not depend on the
number $q$ of $E$-queries made by $A$ or the block length $\blocklen$ of the
cipher. This reflects the reality.  In fact our result is essentially tight;
more precisely, the bound above is tight up to constant factors as long as $q$
is not too tiny. This is established by \thref{th-att-double-des} where we show
that an appropriate adaptation of the standard meet in the middle attack
enables an adversary to obtain an advantage close to that of the upper bound.

The rest of this section will be devoted to a proof of
\thref{th-double-des}.
We fix an adversary $A$ who makes at most $q$ $\E$ queries and at most $t$
$\F/\F^{-1}$ queries. We want to show that
$\Adv_A(\dfo,\keylen,\blocklen)\leq
t^2/2^{2\keylen}$.  We will first introduce some terminology.

\subsection{Preliminaries}

\headingg{The probability spaces.} We consider two ``games.'' Each consists of
running the adversary with its oracles chosen according to some probability
space. Probability Space~1 is that of the experiment defining $\SuccIdeal_A(
\dfo,\keylen,\blocklen)$. Namely, the underlying experiment is:
\begin{newmath}
  F\gets\BC{\keylen}{\blocklen}\next
  k_1^*\gets\bits^{\keylen}\next
  k_2^*\gets\bits^{\keylen}\next E\gets F_{k_1^*}\circ F_{k_2^*} \;,
\end{newmath}%
and Game~1 is to just run $A^{E,F,F^{-1}}$ and reply to its oracle queries
according to the functions $E,F,F^{-1}$ chosen by the experiment.  Now, the
experiment defining Probability Space~2 is
\begin{newmath}
  F\gets\BC{\keylen}{\blocklen}\next
  k_1^*\gets\bits^{\keylen}\next
  k_2^*\gets\bits^{\keylen}\next E\gets \PERM{\blocklen} \;.
\end{newmath}%
In Game~2, we just run $A^{E,F,F^{-1}}$ and reply to its oracle queries
according to the functions $E,F,F^{-1}$ chosen by the experiment. Notice that
in so doing, we completely ignore the two keys $k_1^*,k_2^*$; the responses to
oracle queries do not depend on these at all. Thus, the output of $A$ in Game~2
is exactly that in the experiment defining $\SuccIdeal_A(\keylen,\blocklen)$.
The extra keys we have created will be used only in the analysis. We let
$\Probb{1} {\cdot}$ denote the probability under Probability Space~1, and
$\Probb{2}{\cdot}$ that under Probability Space~2.

\heading{Quantities involved.} Since we are not limiting the computing power of
the adversary, we may, without loss of generality, regard it as deterministic.
We may also assume it makes exactly $q$ $\E$ queries and exactly $t$
$\F/\F^{-1}$ queries, and that no query is ever repeated.  When the oracles
$E,F,F^{-1}$ are fixed, the sequence of queries by $A$ and responses by the
oracles is determined. We view it as a game in which the adversary and the
oracles alternate moves; one query followed by a response is a round, so each
round has two moves, the first by the adversary, the second by the
oracles. There are $q+t$ rounds. We will be referring to the following
quantities:
\medskip
\renewcommand{\arraystretch}{1.2}

\begin{tabular}{rcp{5in}}
$\Moves$ & = & The set $\sset{0,1,\ldots,2(q+t)}$ whose members will be used
to index moves of the game. \\
$\OddMoves$ & = & The set of odd numbers in $\Moves$, corresponding to
question moves. \\
$\EvenMoves$ & = & The set of even numbers in $\Moves$, corresponding to 
   reply moves.
\end{tabular}  
\medskip

\noindent
It is technically convenient to include $0$ in these sets even though there is
no $0$-th round or move. Furthermore we use the following notation:
\medskip

\begin{tabular}{rcp{4.5in}}
$q_i$ & : & For $i\in\OddMoves$, the query in the $i$-th move. It is of
the form $(x,*)$, $(k,x,*)$, or $(k,*,y)$ which are queries to $E$, $F$, 
and $F^{-1}$, respectively. \\
$r_i$ & : & For $i\in\EvenMoves$, the reply in the $i$-th move. For $i>0$
it is $(x,E(x))$, $(k,x,F_k(x))$, or $(k,F_k^{-1}(y),y)$, corresponding,
respectively, to the query $q_{i-1}$; for $i=0$ it is the empty string. \\
$\vview_i(A^{E,F,F^{-1}})$ & : & For $i\in \Moves$, the view of the
adversary
after $i$ moves; this is $q_{1}r_2\ldots q_{i-1}r_{i}$ if $i>0$ is even;
$q_{1}r_1\ldots r_{i-1}q_{i}$ if $i$ is odd; and the empty string if $i=0$
\\
$\vview(A^{E,F, F^{-1}})$ & : & $\vview_{2(q+t)}(A^{E,F,F^{-1}})$.
\end{tabular}
\medskip

\noindent
Note the adversary's output bit is some deterministic function of the last
view.  We call the keys $(k_1^*,k_2^*)$ chosen in the games the \textit{crucial
key pair}. Our analysis will focus on whether or not this key pair is
``eliminated'' by a current view, and what is its distribution {from} the point
of view of $A$ if not.  So let $v_i$ represent a possible view after $i$ moves
of the game.  We consider two sets of key pairs, the ``seen key pairs'' (SKP)
and the ``remaining key pair'' (RKP):
\medskip

\begin{tabular}{rcp{5in}}
$\SeenKeyPairs(v_i)$ & : & A key pair
$k_1,k_2$ is in $\SeenKeyPairs(v_i)$ if there are two queries $q$ and $q'$
in
$v_i$ such that $q$ is an $F$-query or $F^{-1}$ query with key $k_1$ (i.e.,
a
query of the form $(k_1,x,*)$ or $(k_1,*,y)$, respectively), and $q'$ is an
$F$-query or $F^{-1}$ query with key $k_2$ (i.e., a query of the form
$(k_2,x,*)$ or $(k_2,*,y)$, respectively). \\
$\RemKeyPairs(v_i)$ & = & $(\bits^{\keylen}\cross \bits^{\keylen})-
   \SeenKeyPairs(v_i)$ 
\end{tabular}
\medskip

\noindent
Note that $\SeenKeyPairs(v_i)$ depends only on the queries in $v_i$ and not on
the replies. That is, $\SeenKeyPairs(v_{i}) = \SeenKeyPairs(v_{i+1})$ for $i\in
\OddMoves$.  If $A$ knows that $F_{k_2}(x)=y$ and $F_{k_1}(y)=z$ and has also
made the $E$ query $x$ then it can with high probability eliminate $(k_1,k_2)$
as a candidate for the crucial key pair. Intuitively, we might think of the key
pairs $(k_1,k_2)\in\SeenKeyPairs(v)$ as being ``eliminated''. (Of course, they
might not be eliminated, but we can't be sure, so we count them out.) Thus
$\RemKeyPairs(v_i)$ captures the set of remaining key pairs associated to any
view. These are the key pairs $(k_1,k_2)$ so that at least one of them has not
been in either an $F$ or an $F^{-1}$ query.  Note the key pair is \textit{not}
considered ``eliminated'' if one of its components has been in a $F/F^{-1}$
query: \textit{both} have to have been in such queries to ``eliminate'' the
pair.

The current view $v_i$ contains some number of $F$ or $F^{-1}$ queries on a
particular key $k$. This effectively ``opens up'' the corresponding spots in
row $k$ of the $F$ table, in the sense that in the randomly chosen $F$ table,
these entries become known to the adversary.  Similarly for $E$-queries. We let
\medskip

\begin{tabular}{rcp{5in}}
 $\Fqueries(v_{i},k)$ & = & The set of all $y$ such that there are 
 responses in $v_{i}$ of the form $(k,x,y)$.  \\
 $\Equeries(v_{i})$ & = & The set of all $y$ such that there are 
   responses in $v_{i}$ of the form $(x,y)$.  
\end{tabular}
\medskip

%
%

\heading{The random variables.} Under the random choice of $E,F,F^{-1}$ made in
the probability spaces~1 and~2, the above discussed quantities become random
variables. Here are some random variables we will need to refer to explicitly:
\medskip

\begin{tabular}{rcp{5in}}
$\rvQ_i$ & : & Takes value $q_i$, the $i$-th query, for $i\in\OddMoves$. \\
$\rvR_i$ & : & Takes value $r_i$, the $i$-th reply, for $i\in\EvenMoves$. \\
$\rvT_i$ & : & Equals $\rvQ_i$ if $i$ is odd and $\rvR_i$ if $i$ is even. \\
$\rvview_i$ & : & Takes value $\vview_i(A^{E,F,F^{-1}})$, for $i\in\Moves$.
\\
$\rvview$ & : & Takes value $\vview(A^{E,F,F^{-1}})$. \\
$\rvU_{i,j}$ & : & Equals $\rvT_i\ldots\rvT_j$ \\
\end{tabular}  
\medskip

\heading{The bad event.} We also define a central event: 
\medskip

\begin{tabular}{rcp{5in}}
$\bad_i$ & : & For $i\in\Moves$, event $\bad_i$ is said to happen if the 
crucial key pair $(\rvk^*_1,\rvk^*_2)$ is seen, that is, $(\rvk^*_1,
\rvk^*_2)\in\SeenKeyPairs(\rvview_i)$.
\end{tabular} 
\medskip

\noindent In other words, the crucial key pair is ``eliminated''.  Whether a
particular key pair has been seen only depends on the queries of $A$ and
thus
$\bad_{i} =\bad_{i+1}$ for $i\in \OddMoves$.  We let $\bad$ be
$\bad_{2(q+t)}$,
meaning it captures whether the bad event happened at the end of the game.

\subsection{Proof outline}

A very rough cut at the idea of the analysis is that as long as $\bad$ has not
happened in probability space~1, the answers coming back to oracle queries
there ``look random'' and so probability space~1 looks like probability
space~2. We can then bound the advantage by the probability of the bad event.

This is overly simplistic. It is also incorrect. One should first note that
even if the bad event fails to happen in game~1, that game will not look like
game~2; there are events that have probability one in the latter and zero in
the former. In fact, we need to condition on the bad event not happening in
\textit{both} probability spaces.

We will show that the conditional probability of a particular view given that
$\bad$ has not occurred is the same in the two games.  To show this we will be
forced to show something stronger as stated in the lemma below.
\begin{lemma}\label{lm-views-same} Let $i\in\Moves$ and let $v_i$ be a
possible
view of the adversary after the $i$-th move. Then for all $0\leq s \leq 
2(q+t)-i$,
\begin{eqnarray*}
    \CondProbb{1}{\rvview_i=v_i}{\notbad_{i+s}} &\eqq &
    \CondProbb{2}{\rvview_i=v_i}{\notbad_{i+s}} .
\end{eqnarray*}
\end{lemma}
The proof of this lemma is postponed until later.
Since the final decision of the adversary depends
only on its view, the distribution of the adversary's
decision is the same in the two games as long as the bad event has not 
happened. Thus, a corollary to the above lemma is
\begin{equation}\label{eq-yes-given-good}
    \CondProbb{1}{A^{\rvE,\rvF,\rvF^{-1}}=1}{\notbad} \eqq 
    \CondProbb{2}{A^{\rvE,\rvF,\rvF^{-1}}=1}{\notbad} \;. \label{eq-fullbad}
\end{equation}
Less obvious is that Lemma~\ref{lm-views-same} will also be needed to
show that the probability of the bad event is the same in both games.
To show this we need to prove something a bit stronger: we need to
show that the equality holds at any stage.  This is stated in the
lemma stated below.
\begin{lemma}\label{lm-bound-bad} For all $i=0,\ldots,2(q+t)$,
\begin{newequation}
\Probb{1}{\bad_i} \eqq \Probb{2}{\bad_i} \;. \label{eq-bound-bad}
\end{newequation}%
\end{lemma}
The proof of this lemma is also postponed until later. 
Lemmas~\ref{lm-views-same} and~\ref{lm-bound-bad} can be used to bound the
advantage of the adversary by the probability of the bad event.
\begin{lemma}\label{lm-bound-adv} $\Adv_A(\dfo,\keylen,\blocklen) \leq 
\Probb{2}{\bad}$.

\begin{proof}[\lemref{lm-bound-adv}] The lemma is shown using the following 
  straightforward calculation.  We suppress the superscripts of
  $A^{\rvE,\rvF,\rvF^{-1}}$ for clarity.
\begin{eqnarray*}
\lefteqn{\Probb{1}{A=1}
     -\Probb{2}{A=1}}\\ & = &
\hspace{10pt}
\CondProbb{1}{A=1}{\notbad}\cdot\Probb{1}{\notbad}
\:-\: \CondProbb{2}{A=1}{\notbad}
\cdot\Probb{2}{\notbad}\\
& & +\:\CondProbb{1}{A=1}{\bad}\cdot\Probb{1}{\bad} \:-\:
\CondProbb{2}{A=1}{\bad}\cdot\Probb{2}{\bad} \\
& = & \hspace{10pt}\left(\CondProbb{1}{A=1}{\notbad}
            -\CondProbb{2}{A=1}{\notbad}\right)
\cdot \Probb{2}{\notbad}) \\
& & + \: \left(\CondProbb{1}{A=1}{\bad}
        -\CondProbb{2}{A=1}{\bad}\right)
            \cdot\Probb{2}{\bad} \\
& = &  \left(\CondProbb{1}{A=1}{\bad}
        -\CondProbb{2}{A=1}{\bad}\right)
            \cdot\Probb{2}{\bad} \;.
\end{eqnarray*}
The second equality follows by Lemma~\ref{lm-bound-bad}.  The last equality
follows by \eqref{eq-fullbad}.
\end{proof}\end{lemma}
Of course, since the probability of the bad event is the same in both
probability spaces we could have bounded the advantage by the probability of
the bad event in probability space 1.  However, calculating the probability
of
the bad event is very easy in probability space 2 as can be seen below.
\begin{lemma}\label{lm-bound-bad-two} $\Probb{2}{\bad} \leqq 
t^2/2^{2\keylen}$.

\begin{proof}[\lemref{lm-bound-bad-two}] This is straightforward, since in 
  Game~2, no information about the keys $(\rvk_1^*,\rvk_2^*)$ is given
  to the adversary. The bad event depends only on the number of $\F$
  and $\F^{-1}$ queries, and in the worst case all the $t$ such
  queries are made to different keys. Then the chance that $\rvk_1^*$
  is in any query is $t/2^{\keylen}$, and the same, independently, for
  $\rvk_2^*$, so the bound holds.
\end{proof}\end{lemma}
Clearly, Lemmas~\ref{lm-bound-adv} and~\ref{lm-bound-bad-two} imply
Theorem~\ref{th-double-des}.  This completes the outline of the proof of
Theorem~\ref{th-double-des}.  To complete the proof we must prove
Lemmas~\ref{lm-views-same} and~\ref{lm-bound-bad}.

To do so we will first need a sequence of three lemmas,
Lemmas~\ref{lm-replies-uniform}, \ref{lm-extension-prob},
and~\ref{lm-replies-given-bad}.  The last of these will be used in the proof of
Lemma~\ref{lm-views-same}.  Lemma~\ref{lm-replies-uniform} will again be used
to prove Lemma~\ref{lm-rem-key-pairs} on the conditional probability of the
crucial key pair.  Lemma~\ref{lm-rem-key-pairs} will then be used with
Lemma~\ref{lm-views-same} to prove Lemma~\ref{lm-bound-bad}.

\subsection{Distribution of replies in the next round} 

In Game~2, given the view $v_i$ at any point, the distribution of the answer to
the next oracle query is, clearly, uniform, over the remaining range; for
example, the answer to an $E$-query is uniform over
$\bits^{\blocklen}-\Equeries(v_i)$.

The first lemma will say this is true for Game~1 too, as long as the bad event
does not happen. However, we will need to say this in a strong sense.  Namely,
fix any key pair that has still not been ``eliminated''. Conditioned on this
being the crucial key pair, as well as on the current view, the distribution of
the answer to the next oracle query is still ``as it should be,'' meaning
uniform over whatever possibilities remain. Note we must show this for all
types of queries: $E,F$ and $F^{-1}$.
\begin{lemma}\label{lm-replies-uniform} Let $j\in\{1,2\}$ and
$i\in\OddMoves$.
Let $v_{i}=q_1r_2\ldots q_{i-2}r_{i-1}q_i$ be a possible view of the
adversary
just before the answer to query $q_i$ is obtained.  For any string $r_{i+1}
\in\bits^{\blocklen}$ and all $(k_1,k_2)\in\RemKeyPairs(v_{i}\concat
r_{i+1})$,
\begin{eqnarray*}
\lefteqn{
  \CondProbb{j}{\rvR_{i+1}=r_{i+1}}{(\rvk_1^*,\rvk_2^*)=(k_1,k_2)
                                  \:\wedge\: \rvview_i = v_{i} 
                                  } \eqq}\\  & &
  \left\{ \begin{array}{ll}
   {\displaystyle \frac{1}{2^n-|\Equeries(v_i)|}}  & 
 \mbox{if  $q_{i}$ is an $\E$-query and $r_{i+1}\not\in\Equeries(v_i)$} \\
   {\displaystyle \frac{1}{2^n-|\Fqueries(k,v_i)|}} & 
             \mbox{if $q_{i}$ is an $F$ or $F^{-1}$ query with key $k$
                    and $r_{i+1}\not\in\Fqueries(k,v_i)$}  \\
   0 & \mbox{otherwise.} \mystrut
\end{array}\right.
\end{eqnarray*}
In particular, the value depends neither on $j$ nor on $(k_1,k_2)$.

\begin{proof}[\lemref{lm-replies-uniform}] This is clear for Game~2, ie.~for
$j=2$. The proof is devoted to showing it also for Game~1, ie.~for $j=1$.

Let $v_{i+1}=v_ir_{i+1}$. We fix a particular key pair $(k_1,k_2)\in
\RemKeyPairs(v_{i+1})$. Assume $\rvview_i=v_i$, and assume $(\rvk_1^*,
\rvk_2^*)=(k_1,k_2)$.  Note this implies that $\notbad_{i+1}$ holds.
Now consider three cases.
\smallskip

{\sl Case 1:\/} $q_{i}$ is an $\E$-query.

We want to show that $\rvR_{i+1}$ is equally likely to be any string not yet
returned as an answer to an $\E$-query.  The danger is that $\F$ or
$\F^{-1}$
queries have been made to at least one of the crucial keys $k_1,k_2$, and
this
is giving some information about $F_{k_1}\circ F_{k_2}$ in addition to that
{from} the $\E$ queries.

However, this won't happen. This can be seen as follows.  We know that
$\notbad_{i+1}$ holds, which means either $k_1$ or $k_2$ has never been in
any
$\F$ or $\F^{-1}$ query of the adversary.  This means that $F_{k_1} \circ
F_{k_2}$, being the composition of two permutations with one random, is
random
{from} the point of view of the adversary. (The probability here is over the
choice of the cipher $F$, which assigns a random permutation to each key.)
Of
course the adversary has partial information about $F_{k_1}\circ F_{k_2}$ in
the form of replies to previous $\E$-queries, but this gives no information
on
the value of any remaining one except that it will not be one already seen.
\smallskip

{\sl Case 2:\/} $q_{i}$ is an $\F$-query.

Let $k$ be the key in the query. If $k\not\in\{k_1,k_2\}$ it is clear that
the
response to the query is randomly distributed over
$\bits^n-\Fqueries(k,v_i)$ just by the random choice of $F$ in the
experiment.  So suppose $k=k_l$ where $l\in \{1,2\}$. Now, the danger is
that
$\E$ queries yielded some information about $F_k$ in addition to the queries
made directly to key $F_k$, so the adversary will have some advantage in
predicting a new value on $F_k$.

However, this will not be true. This can be seen as follows.  We know
$(k_1,k_2)\in\RemKeyPairs(v_{i+1})$, which means that either $k_1$ or $k_2$
has
not been in any $\F$ or $\F^{-1}$ query up to and including the query in
$q_{i}$. Let $\pi = F_{k_1}\circ F_{k_2}$. As the composition of two
permutations, one of which is random, it is random {from} the point of view
of
the adversary. Then $F_k = F_{k_l} = \pi\circ F_{k_2}^{-1}$ if $l=1$ and
$F_k =
F_{k_l} = F_{k_1}^{-1} \circ \pi$ if $l=2$. In either case, $F_k$ is the
composition of two permutations, one of which is random {from} the point of
view of the adversary, and hence the response to an $\F$ query on key $k$
will
return a value distributed uniformly over $\bits^n-\Fqueries(k,v_i)$.

{\sl Case 3:\/} $q_{i}$ is an $\F^{-1}$-query.

The proof that the response to the query is uniformly distributed over
$\bits^n-\Fqueries(k,v_i)$ is similar to the case above.
\end{proof}\end{lemma}

\noindent
The above lemma shows that for a fixed partial conversation $v_i$ where
$i\in\OddMoves$, and fixed pair of keys $k_1,k_2$ such that $\notbad_i$ is
true
(i.e., $(k_1,k_2) \in \RemKeyPairs(v_i)$), all the answers $r_{i+1}$ which
continue to keep the partial conversations {from} being ``bad'' (i.e.,
$(k_1,k_2)
\in \RemKeyPairs(v_ir_{i+1})$), have the same probability in each
probability
space.  We will use this lemma to prove an extension of this.  Namely, for a
fixed partial conversation $v_i$ and fixed pair of keys $k_1,k_2$ such that
$\notbad_i$ is true, all further move sequences which continue to keep the
partial conversations {from} being ``bad'' have the same probability in each
probability space.  We state this formally below.

\begin{lemma}\label{lm-extension-prob} Let $j\in\{1,2\}$. Let $v_i$ be a 
possible view of the adversary after move $i\in \Moves$, and let $1\leq
\ell\leq 2(q+t)-i$. For any possible extension $u_{i+1,i+\ell}$ of $v_i$ by
$\ell$ moves, and for any key pair $(k_1,k_2)\in\RemKeyPairs(v_i\concat
u_{i+1,i+\ell})$,
\[
\CondProbb{j}{\rvU_{i+1,i+\ell}=u_{i+1,i+\ell}}{(\rvk_1^*,\rvk_2^*)=
(k_1,k_2)\:\wedge\: \rvview_i = v_i }
\]
depends neither on $j$ nor on $(k_1,k_2)$. (That is, it depends only on $v_i$
and $u_{i+1,i+\ell}$.)

\begin{proof}[\lemref{lm-extension-prob}] We will prove this by induction on
  $\ell$.  The base case is $\ell=1$. In this case the lemma is clear
  when $i+1=i+\ell$ is odd, because in this case $u_{i+1,i+1}$ is a
  query, which is a function only of $A$ and $v_i$. In the case of
  $i+1=i+\ell$ being even, $u_{i+1,i+1}$ is the response $\rvR_{i+1}$,
  and we can apply Lemma~\ref{lm-replies-uniform}.

Now assume that the lemma is true for $\ell=s$. We want to establish it for
$\ell=s+1$. Again, this is trivial if $i+s+1$ is odd, because then the
extension is a query, uniquely determined given $v_iu_{i+1,i+s}$ and $A$.
So
assume $i+s+1$ is even. Let $u_{i+1,i+s+1} = u_{i+1,i+s}r_{i+s+1}$ and
$v_{i+s}=v_iu_{i+1,i+s}$. We assume that $(k_1,k_2)\in \RemKeyPairs(v_i
u_{i+1,i+s+1})$.  We can write
\begin{eqnarray*}
\lefteqn{\CondProbb{j}{\rvU_{i+1,i+s+1}=u_{i+1,i+s+1}}
 {(\rvk_1^*,\rvk_2^*)=(k_1,k_2) \:\wedge\: \rvview_i = v_i } = } \\
 & &  \CondProbb{j}{\rvR_{i+s+1} = r_{i+s+1} }{
                                (\rvk_1^*,\rvk_2^*)=(k_1,k_2)
                                 \:\wedge\: \rvview_{i+s} = v_{i+s} } \\
  & & \qquad  \cdot 
\CondProbb{j}{\rvU_{i+1,i+s}=u_{i+1,i+s}}{(\rvk_1^*,\rvk_2^*)=(k_1,k_2)
                                  \:\wedge\: \rvview_i = v_i }.
\end{eqnarray*}
The first factor depends neither on $j$ nor on $(k_1,k_2)$ by
Lemma~\ref{lm-replies-uniform}.  The second factor has the same property by
induction.
\end{proof}\end{lemma}
We now use the above lemma to prove a generalization of 
Lemma~\ref{lm-replies-uniform} which we will need subsequently.

\begin{lemma}\label{lm-replies-given-bad} Let $j\in\{1,2\}$ and $i\in
\OddMoves$.  Let $v_{i}=q_1r_2\ldots q_{i-2}r_{i-1}q_i$ be a possible view
of
the adversary just before the answer to query $q_i$ is obtained.  For any
string $r_{i+1} \in\bits^{\blocklen}$, all
$(k_1,k_2)\in\RemKeyPairs(v_{i}\concat r_{i+1})$, and all $0\leq s \leq
2(q+t)-i$,
\[
  \CondProbb{j}{\rvR_{i+1}=r_{i+1}}{
                                  (\rvk_1^*,\rvk_2^*)=(k_1,k_2)
                                  \:\wedge\: \rvview_i = v_i 
                                  \:\wedge\: \notbad_{i+s} }
\]
depends neither on $j$ nor on $k_1,k_2$. (That is, it depends only on $v_i$
and $r_{i+1}$ and $s$.)

\begin{proof}[\lemref{lm-replies-given-bad}]  First suppose $s=0$. The 
  conditioning on $\notbad_i$ is redundant; this event will be true
  because $(k_1,k_2)\in\RemKeyPairs(v_{i} \concat r_{i+1})$. Thus the
  claim is true {from} \lemref{lm-replies-uniform}.

So assume $s\geq 1$.  The probability in the statement of the lemma can be
written as
\[ 
\frac{
 \CondProbb{j}{\rvR_{i+1}=r_{i+1}
                                  \:\wedge\: \notbad_{i+s} }{
                                  (\rvk_1^*,\rvk_2^*)=(k_1,k_2)
                                  \:\wedge\: \rvview_i = v_i  } }{
  \CondProbb{j}{\notbad_{i+s} }{  (\rvk_1^*,\rvk_2^*)=(k_1,k_2)
                               \:\wedge\: \rvview_i = v_i 
                                 }                     }.
\]
The denominator can be written as 
\[
\sum  \CondProbb{j}{\rvU_{i+1,i+s} = u_{i+1,i+s} }{ 
                               (\rvk_1^*,\rvk_2^*)=(k_1,k_2)
                               \:\wedge\: \rvview_i = v_i 
                                 }                    
\]
where the sum is over $u_{i+1,i+s}$ such that $(k_1,k_2)\in
\RemKeyPairs(v_iu_{i+1,i+s})$.  By Lemma~\ref{lm-extension-prob} each term
of
this sum has a value that depends neither on $j$ nor on $(k_1,k_2)$.
The numerator can be written as
\[
\sum  \CondProbb{j}{\rvR_{i+1}\rvU_{i+2,i+s} 
                                 =r_{i+1}u_{i+2,i+s} }{ 
                               (\rvk_1^*,\rvk_2^*)=(k_1,k_2)
                               \:\wedge\: \rvview_i = v_i 
                                 }                    
\]
where the sum is over $u_{i+2,i+s}$ such that $(k_1,k_2)\in \RemKeyPairs(v_i
r_{i+1}u_{i+2,i+s})$.  By Lemma~\ref{lm-extension-prob} each term of this
sum
depends neither on $j$ nor on $(k_1,k_2)$. This completes the proof of the 
lemma.
\end{proof}\end{lemma}

\begin{proof}[\lemref{lm-views-same}] The proof will be by induction on
$i\in
\Moves$.  The base case of the induction is when $i=0$, and in this case the
lemma is trivially true because the view is by definition the empty string.
So
assume the statement of the lemma up to move $i$.  We will prove it for
$i+1$.
Fix an arbitrary $s\geq 0$. 

First consider the case where $i\in\EvenMoves$, meaning the last move in
$v_i$
is a reply. Let $q_{i+1}$ be arbitrary. Then:
\begin{eqnarray*}
\lefteqn{\CondProbb{j}{\rvview_{i+1} = v_iq_{i+1}}{
                  \notbad_{i+1+s}}} \\
 & = & \CondProbb{j}{\rvview_i = v_i}{ 
              \notbad_{i+1+s}} \cdot 
  \CondProbb{j}{\rvQ_{i+1} = q_{i+1}}{
                         \rvview_i = v_i \:\wedge\: 
                      \notbad_{i+1+s}} \;.
\end{eqnarray*}
First, look at the first factor.  Since $s\geq 0$ by assumption, then $s
+1\geq
0$, and therefore the first term is the same for $j=1$ and $2$ by induction.
Next look at the second factor.  $A$'s query is just dependent on $A$ and on
$v_i$, the view so far.  Thus, the probability is the same for both $j=1$
and
$j=2$. (And is equal to $0$ except possibly for one value of $q_{i+1}$.)
Therefore, the product of the two probabilities is equal for $j=1$ and 
$j=2$, for all $s\geq 0$.

Next consider the case where $i\in\OddMoves$, meaning the last move in $v_i$
is a query. Let $r_{i+1}\in\bits^{\blocklen}$ be arbitrary and let $v_{i+1}=
v_ir_{i+1}$. Then:
\begin{eqnarray*}
\lefteqn{\CondProbb{j}{\rvview_{i+1} = v_{i}r_{i+1}}{
              \notbad_{i+1+s}}} \\
 & = & \CondProbb{j}{\rvview_{i} = v_{i}}{ 
             \notbad_{i+1+s}} \cdot 
  \CondProbb{j}{\rvR_{i+1} = r_{i+1}}{\rvview_{i} = v_{i} \:\wedge\: 
                \notbad_{i+1+s}} \;.
\end{eqnarray*}
Consider the first factor. Since $s\geq 0$ by assumption, then $s+1\geq 0$,
and
therefore, by induction, the first term is the same for $j=1$ and $2$.  The 
second factor is equal to:
\[ 
     \sum_{(k_1,k_2)} p_j(k_1,k_2)\cdot q_j(k_1,k_2)
\]
where the sum is over all $(k_1,k_2)\in
\bits^{\keylen}\times\bits^{\keylen}$
and we have set
\begin{eqnarray*}
 p_j(k_1,k_2) & = &
\CondProbb{j}{\rvR_{i+1} = r_{i+1}}{(\rvk_1^*,\rvk_2^*)=(k_1,k_2) \:\wedge\:
     \rvview_{i} = v_{i} \:\wedge\: \notbad_{i+1+s}} \\
 q_j(k_1,k_2) & = & 
  \CondProbb{j}{(\rvk_1^*,\rvk_2^*)=(k_1,k_2)}{
                    \rvview_{i} = v_{i} \:\wedge\:
                  \notbad_{i+1+s}}
\end{eqnarray*}
We start by examining the first factor, namely $p_j(k_1,k_2)$. By
Lemma~\ref{lm-replies-given-bad}, for all $(k_1,k_2)\notin
\SeenKeyPairs(v_{i+1})$, this probability is the same for both $j=1$ and
$2$,
and independent of $(k_1,k_2)$. Call this value $p$. On the other hand for
$(k_1,k_2)\in \SeenKeyPairs(v_{i+1})$ we have $p_j(k_1,k_2)=0$ because
of the conditioning on $\notbad_{i+1+s}$. Thus the above sum reduces to
\[
   p\cdot \sum_{(k_1,k_2)} q_j(k_1,k_2)
\]
where the sum is over all $(k_1,k_2)\in \RemKeyPairs(v_{i+1})$.  We claim
that
this range is over all the nonzero values of the probability and thus the
sum
is equal to 1.  To see this, note that $q_j(k_1,k_2)$ is equal to 0 for
$(k_1,k_2) \in \SeenKeyPairs(v_{i+1})$.  This completes the induction and
the
proof of Lemma~\ref{lm-views-same}.
\end{proof}

\noindent
The remaining task is to prove Lemma~\ref{lm-bound-bad} which states that
the
probability that the bad event occurs is the same in both probability
spaces.
To do so we will first prove the following lemma about the distribution of
keys.  The proof of this lemma will use Lemma~\ref{lm-views-same} which,
recall, states that the probability of a given query and response (which are
not bad) for a fixed partial view and a fixed pair of keys (which are not
bad)
is the same in both probability spaces.

\subsection{Equi-probability of unseen keys} 

A crucial lemma is that in Game~1, as long as the bad event has not
happened,
if adversary has a particular view, then any ``un-eliminated'' key pair is
equally likely to be the crucial key pair. Without this, it might be that
the
adversary's chance of hitting the crucial key is better in Game~1 (given the
bad event fails) than in Game~2 (given the bad event fails).  To simplify
notation, for $j\in\{1,2\}$ and $v_i$ let
\begin{newmath}
  \Probb{j,v_i}{\cdot} \eqq \CondProbb{j}{\cdot}{
                 \rvview_i=v_i\evand\notbad_i}\;.
\end{newmath}%

\begin{lemma}\label{lm-rem-key-pairs} Let $j\in\{1,2\}$. Let $v_{i}$ be a 
possible view of the adversary after move $i\in\Moves$. Let $(k_1,k_2)\in
\RemKeyPairs(v_{i})$. Then
\begin{newmath}
  \Probb{j,v_{i}}{(\rvk_1^*,\rvk_2^*)=(k_1,k_2)} \eqq \frac{1}{
        |\RemKeyPairs(v_{i})|}\;.
\end{newmath}%

\begin{proof}[\lemref{lm-rem-key-pairs}] This is clear in Game~2, ie.~for 
$j=2$.  The proof is devoted to showing the claim in Game~1, ie.~for $j=1$.
The proof will be by induction on the move number $i\in\Moves$.  The base
case is $i=0$. In this case no queries have been made so the adversary has
no
information about $(\rvk_1^*,\rvk_2^*)$, and all possible pairs of keys
remain
equally likely, so the claim is true.  So, assume the lemma statement is
true
up to move $i\in\Moves$ where $i< 2(q+t)$.  We will prove it for $i+1$.

Let $v_{i+1}=v_i\tau$ where $\tau=q_{i+1}$ is a query if $i$ is even and
$\tau=r_{i+1}$ is a reply if $i$ is odd. Assume $(k_1,k_2)$ is some key 
pair. Consider the quantity
\begin{newequation}
  \Probb{1,v_{i+1}}{(\rvk_1^*,\rvk_2^*)=(k_1,k_2)} \;.
  \label{eq-rem-key-pairs-ext}
\end{newequation}%

{\sl Claim 1:\/} The quantity of \eqref{eq-rem-key-pairs-ext} is zero if
$(k_1,k_2)\not\in\RemKeyPairs(v_{i+1})$.

{\sl Proof of Claim 1:\/} This is because $\Probb{1,v_{i+1}}{\cdot}$
conditions on $\notbad_{i+1}$, meaning we know $\bad_{i+1}$ did not
happen.~$\Box$
\smallskip

{\sl Claim 2:\/} Let $(k_1,k_2)$ be any key pair in $\RemKeyPairs(v_{i+1})$.
Then the quantity of \eqref{eq-rem-key-pairs-ext} has a value that depends
only
on $v_{i+1}$, and not on $(k_1,k_2)$.

We will prove Claim~2 below. The two claims together imply that the only
possibility is that for all $(k_1,k_2)\in\RemKeyPairs(v_{i+1})$,
\begin{newmath}
  \Probb{1,v_{i+1}}{(\rvk_1^*,\rvk_2^*)=(k_1,k_2)} \eqq
   \frac{1}{|\RemKeyPairs(v_{i+1})|} \;.
\end{newmath}%
Thus the induction would be completed.
\smallskip

{\sl Proof of Claim 2:\/} Recall $\rvT_{i+1}=\rvQ_{i+1}$ if $i$ is even and
$\rvT_{i+1}=\rvR_{i+1}$ if $i$ is odd. Expand the quantity of 
\eqref{eq-rem-key-pairs-ext}:
\begin{newmath}
\Probb{1,v_{i+1}}{(\rvk_1^*,\rvk_2^*)=(k_1,k_2)} \eqq
\CondProbb{1,v_{i}}{(\rvk_1^*,\rvk_2^*)=(k_1,k_2)}
               {\rvT_{i+1}=\tau\evand\notbad_{i+1}}\\
\end{newmath}%
and then apply Bayes rule to get:
\begin{newmath}%
\CondProbb{1,v_{i}}{\rvT_{i+1}=\tau\evand\notbad_{i+1}}
               {(\rvk_1^*,\rvk_2^*)=(k_1,k_2)}
\cdot \frac{\Probb{1,v_{i}}{(\rvk_1^*,\rvk_2^*)=(k_1,k_2)}}
           {\Probb{1,v_{i}}{\rvT_{i+1}=\tau\evand\notbad_{i+1}}} \;.
\end{newmath}%
We want to argue this quantity does not depend on $(k_1,k_2)$. Look first at
the fraction. The value of the numerator is given by the induction
hypothesis
and in particular does not depend on $(k_1,k_2)$. The value of the
denominator
obviously does not depend on $(k_1,k_2)$ since that quantity appears nowhere
in it. Thus, what is left is to show that
\begin{newequation}
\CondProbb{1,v_{i}}{\rvT_{i+1}=\tau\evand\notbad_{i+1}}
               {(\rvk_1^*,\rvk_2^*)=(k_1,k_2)}
 \label{eq-rem-key-pairs-last}
\end{newequation}%
does not depend on $(k_1,k_2)$. 

Observe that the conjunction of the event $\notbad_{i+1}$ in the probability
of
\eqref{eq-rem-key-pairs-last} is redundant: since we are conditioning on
$(\rvk_1^*,\rvk_2^*)=(k_1,k_2)$, and we know that
$(k_1,k_2)\in\RemKeyPairs(v_{i+1})$, the conditioning already tells us that
$\notbad_{i+1}$ will hold. In other words,
\begin{newmath}
\CondProbb{1,v_i}{\rvT_{i+1}=\tau\evand\notbad_{i+1}}
               {(\rvk_1^*,\rvk_2^*)=(k_1,k_2)}
 \eqq 
\CondProbb{1,v_{i}}{\rvT_{i+1}=\tau}{(\rvk_1^*,\rvk_2^*)=(k_1,k_2)} \;.
\end{newmath}%
Now we consider separately the case where $i+1$ is odd (meaning $\rvT_{i+1}
=\rvQ_{i+1}$ and $\tau=q_{i+1}$) and the case where $i+1$ is even (meaning
$\rvT_{i+1}=\rvR_{i+1}$ and $\tau=r_{i+1}$). In the first case, note that
the
query made is determined only by ($A$ and) the view $v_i$, so the
probability
in question does not depend on $(k_1,k_2)$. In the second case, we can apply
\lemref{lm-replies-uniform} which gives the value of the above quantity for
each $(k_1,k_2)\in \RemKeyPairs(v_{i+1})$, and, as we see {from}
\lemref{lm-replies-uniform}, that value does not depend on $(k_1,k_2)$. This
completes the proof of Claim~2.~$\Box$
\end{proof}\end{lemma}
Using the above lemma we can now prove Lemma~\ref{lm-bound-bad} which
(recall) states that $\probb{1}{\bad_i} = \probb{2}{\bad_i}$ for all 
$i\in \Moves$.

\begin{proof}[\lemref{lm-bound-bad}] The proof is by induction on
$i\in\Moves$.
The base case is when $i=0$. In this case, the current view $v$ of the 
adversary, in either game, is empty, so that
$\SeenKeyPairs(v)=\emptyset$. Thus, both probabilities are zero.

So, assume the lemma statement is true up to move $i\in\Moves$ where $i <
2(q+t)$.  We will prove it for $i+1$, namely we will show that
\begin{newequation}
\Probb{1}{\bad_{i+1}} \eqq \Probb{2}{\bad_{i+1}} 
\;. \label{eq-bound-bad-next}
\end{newequation}%
We first consider the case where $i+1$ is even, meaning the last move in
$v_i$
is a query. We have
\begin{newmath}
\Probb{j}{\bad_{i+i}} \eqq \Probb{j}{\bad_{i}} \:+\:
   \CondProbb{j}{\bad_{i+1}}{\notbad_{i}} \;.
\end{newmath}%
The first term is equal for $j=1$ and $2$ by induction, and $\CondProbb{j}
{\bad_{i+1}}{\notbad_{i}} = 0$ because $i+1$ is even.

To complete the induction we need to prove \eqref{eq-bound-bad-next} for the
case where $i+1$ is odd, meaning the last move in $v_i$ is a reply. Let
$j\in\{1,2\}$. We can write
\begin{newmath}
\Probb{j}{\bad_{i+1}} \eqq \Probb{j}{\bad_{i}} \:+\:
   \CondProbb{j}{\bad_{i+1}}{\notbad_{i}} \;.
\end{newmath}%
The first term is independent of $j$ by the induction hypothesis. We will
now
argue that the second term is also independent of $j$. By conditioning we
can write the second term as
\begin{eqnarray*}
\CondProbb{j}{\bad_{i+1}}{\notbad_{i}} & = &
\sum_{v_i\in V_j}\: 
\CondProbb{j}{\bad_{i+1}}{\notbad_i\evand\rvview_i=v_i}\cdot
         \CondProbb{j}{\rvview_i=v_i}{\notbad_i} \\
 & = & \sum_{v_i\in V_j}\: \underbrace{
\underbrace{\Probb{j,v_i}{\bad_{i+1}}}_{\textrm{\small first term}}
\cdot 
\underbrace{\CondProbb{j}{\rvview_i=v_i}{\notbad_i}}_{\textrm{\small second
term}}
}_{\textrm{\small product term associated to $v_i$}} \;,
\end{eqnarray*}
where $V_j=\set{v_i}{\CondProbb{j}{\rvview_i=v_i}{\notbad_i}>0}$ is the set
of possible views after move $i$ in Game~$j$.

Let us first observe that $V_1=V_2$, namely the set of views $v_i$ for which
the second term of the ``product term associated to $v_i$'' is positive is
the
same in both games. This is true by \lemref{lm-views-same}, which tells us
that
$\CondProbb{j}{\rvview_i =v_i}{\notbad_i}$ does not depend on $j$ and hence
in
particular the values of $v$ for which it is zero are the same for $j=1$ and
$j=2$.

Now let us set $V=V_1=V_2$ and compare the sums, term by term, in the cases
$j=1$ and $j=2$.  Fix a particular string $v_i\in V$ and focus on the
``product
term associated to $v_i$.''  The second term in it is independent of $j$ by
\lemref{lm-views-same}. We will show the same is true for the first term,
which
will complete the proof. (One needs to be a little careful.  The first term
is
not well defined for just any $v$, only for $v_i\in V_j$.  That's why it was
important, first, to restrict attention to these $v_i$ values, and, second,
to
make sure that $V_1=V_2$, since otherwise we would not be sure that we have
shown equality for every term in the two sums.)

So the remaining task is to consider $\CondProbb{j}{\bad_{i+1}}{\notbad_i
\evand\rvview_i=v_i}$ for $v_i\in V$ and show it does not depend on $j$.
First
note that $\RemKeyPairs(v_i)\neq \emptyset$, because, $\RemKeyPairs(v_i)=
\emptyset$ would imply $\CondProbb{j}{\rvview_i=v_i}{\notbad_i}=0$, and we
have
assumed the last to not be true.  

Since the view $v_i$ and the adversary are fixed, the next query $q_{i+1}$
is
uniquely determined.  Let
\begin{newmath}
\NewKeyPairs(v_i,q_{i+1}) = \RemKeyPairs(v_i) - \RemKeyPairs(v_i\|q_{i+1})
\end{newmath}%
be the set of ``new key pairs'' that are ``seen'' by the $(i+1)$-th query. 
(This set is empty if the latter is an $E$-query.  It is also empty if it is
an
$F$ or $F^{-1}$ query with key with which $A$ has already queried.  If it is
an
$F$ or $F^{-1}$ query with key $k$ with which $A$ has not queried, then the
set
consists of pairs $(k,k')$ and $(k',k)$ where $k'$ is any other key with
which
$A$ has queried $F$ or $F^{-1}$.) We claim that
\begin{newequation}
\CondProbb{j}{\bad_{i+1}}{\notbad_i\evand\rvview_i=v_i}
 \eqq \frac{|\NewKeyPairs(v_i,q_{i+1})|}
                                    {|\RemKeyPairs(v_i)|} \;,
   \label{eq-bound-bad-remaining}
\end{newequation}%
for both $j=1$ and $j=2$.  Note the fraction is well defined, in that the
denominator is not zero, because $\RemKeyPairs(v_i)$ is non-empty.

\eqref{eq-bound-bad-remaining} follows {from} \lemref{lm-rem-key-pairs}.
This
tells us that {from} the point of view of the adversary, all remaining key
pairs remain equally likely, in either game.
\end{proof}

\appendix 

\section{Best attack: Meet in the middle}\label{se-att-df}

In this section we will show the following:
\begin{lemma}\label{lm-att-double-des} For any $\keylen,\blocklen \geq 1$,
any $1\leq s\leq q\leq 2^{\blocklen-1}$, and any $t\geq 2s$, there is an 
adversary $A$ such that
\begin{newmath}
   \Adv_A(\dfo,\keylen,\blocklen) \geqq \frac{t^2}{4s^2}\cdot
      \left(\frac{1}{2^{2\keylen}}- \frac{1}{2^{s(n-1)}}\right) \;.
\end{newmath}%
\end{lemma}
We can now optimize the value of $s$ and obtain the following theorem which
says that the bound of \thref{th-double-des} is essentially tight:
\begin{theorem}\label{th-att-double-des} For any $\keylen,\blocklen\geq 1$,
let $s = \lceil (2\keylen+1)/(\blocklen-1)\rceil$. Then for any $t\geq 2s$
and $s \leq q \leq 2^{n-1}$ it is the case that
\begin{newmath}
   \SecIdeal(\dfo,\keylen,\blocklen,q,t) \geqq \frac{1}{8s^2}
                        \frac{t^2}{2^{2\keylen}} \;.
\end{newmath}%

\begin{proof} The choice of $s$ guarantees that $2^{2\keylen+1}\leq 
2^{s(n-1)}$. This means that
\begin{newmath}
\frac{1}{2^{2\keylen}}- \frac{1}{2^{s(n-1)}} \geqq \frac{1}{2}
                       \frac{1}{2^{2\keylen}} \;.
\end{newmath}%
Now apply \lemref{lm-att-double-des}.
\end{proof}\end{theorem}
Notice that for typical block cipher parameters $\keylen,\blocklen$, the value
of $s$ is very small. For example, for the $\DES$ parameters $\keylen=56$ and
$\blocklen=64$ we have $s=\lceil 113/63\rceil = 2$. Thus the above lower bound
of \thref{th-att-double-des} is in practice close to the upper bound of
\thref{th-double-des}.

\begin{proof}[\lemref{lm-att-double-des}] The proof is by presenting an 
adversary $A$ who achieves the claimed advantage.  The adversary $A$
plays a version of the meet-in-the-middle attack, but we need to
adapt it slightly and then analyze it in our framework. It is
convenient to let $[N]=\{1,2,\ldots,N\}$ for any integer $N\geq 1$.
The adversary proceeds as follows:
\begin{tabbing}
123\=123\=123\=123\=123\=\kill
\textbf{For} $j=1,\ldots,s$ \textbf{do} \\
\> Let $x_j\in\bits^n$ be the $j$-th string in lexicographic order \\
\> Compute $y_j=E(x_j)$ \\
\textbf{Endfor} \\
Choose two disjoint sets $K_1=\set{k_{1,i}} {i\in [t/2s]}$ and $K_2= 
\set{k_{2,i}}{i\in [t/2s]}$ of $\keylen$-bit keys, \\
each set being of size $t/2s$. (These might be chosen at random, 
but not necessarily).\\
\textbf{For} $i=1,\ldots,t/2s$ \textbf{do} \\ 
\> \textbf{For} $j=1,\ldots,s$ \textbf{do} Compute $u_{i,j}=F(k_{1,i},x_j)$
and
    $v_{i,j}=F^{-1}(k_{2,i},y_j)$ \textbf{Endfor} \\
\> Let $u_i=(u_{i,1},\ldots,u_{i,s})$ and $v_i=(v_{i,1},\ldots,v_{i,s})$ \\
\textbf{Endfor} \\
Let $C=\set{(a,b)\in [t/2s]\times [t/2s]}{u_a=v_b}$ \\
\textbf{If} $C\neq\emptyset$ \textbf{then return} $1$ \textbf{else return}
$0$
\end{tabbing}
We now analyze this attack. The first claim is that the cost is as
claimed, meaning $A$ makes at most $q$ $E$-queries and at most $t$
$F/F^{-1}$ queries.  The first is true because $s\leq q$ by
assumption. The second is true because the number of calls to
$F/F^{-1}$ above is $2[(t/2s)s]=t$.  We now want to lower bound
\begin{newmath}
 \Adv_A(\dfo,\keylen,\blocklen) \eqq
\SuccIdeal_A(\dfo,\keylen,\blocklen) -
               \SuccIdeal_A(\keylen,\blocklen)  \;. 
\end{newmath}%
We will lower bound the first term and upper bound the second. Let
$\Prob{\cdot}$ denote the probability in the experiment underlying the
definition of $\SuccIdeal_A(\dfo,\keylen,\blocklen)$, and let
$k_1^*k_2^*$ denote the randomly chosen $2\keylen$ bit key in this
experiment.  Observe that if $k_1^*\in K_1$ and $k_2^*\in K_2$ then
$C$ is definitely non-empty.  So
\begin{newmath}
    \SuccIdeal_A(\dfo,\keylen,\blocklen) \geqq \Prob{k_1^*\in K_1\mbox{ and
}
    k_2^*\in K_2} \eqq \left(\frac{t/2s}{2^{\keylen}}\right)^2 \eqq
                  \frac{1}{4s^2}\frac{t^2}{2^{2\keylen}} \;.
\end{newmath}%
Now let $\Prob{\cdot}$ denote the probability in the experiment
underlying the definition of $\SuccIdeal_A(\keylen,\blocklen)$, and
observe that 
\begin{newmath}
 \SuccIdeal_A(\keylen,\blocklen)  \eqq \Prob{C\neq\emptyset} \;.
\end{newmath}%
For a fixed $a,b\in [t/2s]$ we have
\begin{newmath}
 \Prob{u_a=v_b} \eqq 
          \prod_{j=1}^{s} \frac{1}{N-j-1} \leqq 
           \left(\frac{1}{2^{n-1}}\right)^s \;.
\end{newmath}%
The last inequality here is by the assumption that $s\leq 2^{n-1}$. By the
union bound we have 
\begin{newmath}
    \Prob{C\neq\emptyset} \leqq \frac{t^2}{4s^2}\cdot
\frac{1}{2^{s(n-1)}}\;.
\end{newmath}%
This completes the proof.
\end{proof}

\section{Analysis of the two-key triple cipher}\label{se-triple}

The two-key triple cipher (namely, the construction underlying two-key
triple DES) was defined in \secref{sec-defs}. The same upper bound on
the advantage of any adversary $A$ attacking this cipher can be shown
as for the double cipher:
\begin{theorem}\label{th-triple-des} For any $\keylen,\blocklen,q,t \geq 1$ it
is the case that
\begin{newmath}
 \SecIdeal(\tfo,\keylen,\blocklen,q,t)\leqq \frac{t^2}{2^{2\keylen}} \;.
\end{newmath}%
\end{theorem}
Unlike the case of the double cipher, however, this bound is not tight, and
we believe it can be improved by a better analysis.

The proof of the theorem is obtained by adapting the proof of
\thref{th-double-des}.  We will use essentially the same setup; we start by
giving some new definitions and then continue by showing the necessary
modifications for the proof in Section~\ref{se-double-des} so that it works
also in the case of operator $\tfo$. 
%
%
large

\heading{Games, setup, random variables and event $\bad_i$.}  The experiment
underlying Game 2 is the same as for the proof of \thref{th-double-des}. The
experiment underlying Game 1 is now the following:
\begin{newmath}
  F\gets\BC{\keylen}{\blocklen}\next
  k_1^*\gets\bits^{\keylen}\next
  k_2^*\gets\bits^{\keylen}\next E\gets F_{k_1^*}\circ F^{-1}_{k_2^*}\circ
F_{k_1^*} \;,
\end{newmath}%
and the game is to just run $A^{E,F,F^{-1}}$ and reply to its oracle queries
according to the functions $E,F,F^{-1}$ chosen by the experiment.  The setup
and the random variables are defined exactly in the same way as before, with
the understanding that when we mention $E$-queries, in Game~1, we refer to a
query to the cipher $F_{k_1^*}\circ F^{-1}_{k_2^*}\circ F_{k_1^*}$.  Event
$\bad_i$ is formally defined exactly as before.

\heading{Analysis.}  We observe that almost all lemmas in our previous
analysis
do not significantly depend on the construction we are analyzing.  More
precisely, we see that all lemmas but Lemma~\ref{lm-replies-uniform} require
no
modification for both the statement and the proof to hold also in the case
of
the construction $\tfo$.  So it remains to modify
Lemma~\ref{lm-replies-uniform} so that it works also in the current case.
Recall that such lemma is trying to show that the distribution of the next
reply is independent of which game the adversary is in, and also of a fixed
un-eliminated key pair. However, this can in fact be seen to still be true,
because we are still looking at compositions of random permutations with one
unknown. We omit the details.

\heading{Lower bound.} The standard meet in the middle attack for triple DES
\cite{DiHe,MeHe} can be put and analyzed in our model analogously to the way we
did it above for the double cipher. The analysis indicates that our upper bound
for the two-key triple cipher is tight (up to a constant factor) when $q\approx
t$, but not tight in general. We do not include the details of this analysis.

\end{document}